\documentclass[preprint,5p]{elsarticle}
\makeatletter
\def\ps@pprintTitle{%
 \let\@oddhead\@empty
 \let\@evenhead\@empty
 }
\makeatother

\usepackage[utf8]{inputenc}
\usepackage{graphicx}
\usepackage{slashed}
\usepackage{color}
\usepackage{amsmath}
\usepackage{amssymb}
\usepackage{tabularx}

\usepackage{ulem}

\usepackage{tikz} 
\usetikzlibrary{arrows} 
\usetikzlibrary{shapes,arrows,positioning,automata,backgrounds,calc,er,patterns}
\usepackage{tikz-feynman}

\usepackage{color} 

\definecolor{magenta}{RGB}{255,0,255} 

\definecolor{mygreen}{RGB}{0,128,0} 
 
\definecolor{orange}{RGB}{255,165,0} 
  
\usepackage{soul}

\newcommand{\op}{\mathcal{O}}

\newcommand{\mF}{\mathcal{F}}

\newcommand{\mO}{\mathcal{O}}

\begin{document}
%%%%%%%%%%%%%%%%%%%
% we use this for tables sometimes but it's not necessary
\newcolumntype{C}[1]{>{\hsize=#1\hsize\centering\arraybackslash}X}

\begin{frontmatter}

%%%%%%%%%%%%%%%%%%%
\title{SMEFT is falsifiable through multi-Higgs measurements \\
(even in the absence of new light particles) } 

%%%%%%%%%%%%%%%%%%%

\author[inst2]{Raquel G\'omez-Ambrosio}
\affiliation[inst2]{ 
            organization={Dipartimento di Fisica, Univ. di Torino, and INFN, Sezione di Torino},
    addressline={Via P. Giuria 1}, 
    city={Torino},
    postcode={10125},
    country={Italy.}
}

\author[inst1]{Felipe J. Llanes-Estrada}

\affiliation[inst1]{organization={Univ. Complutense de Madrid, Dept. Fisica Teorica and IPARCOS},
            addressline={Plaza de las Ciencias 1}, 
            city={Madrid},
            postcode={28040},
            country={Spain.}}

\author[inst1]{Alexandre Salas-Bern\'ardez}

\author[inst1]{Juan J. Sanz-Cillero}

%%%%%%%%%%%%%%%%%%%%%%%%%%%%%
\date{\today}
%%%%%%%%%%%%%%%%%%%%%%%%%%%%%%%%%%%%%%%%%%%%%%%%%%%%%%%%%%%%%%%%%%%%
\begin{abstract}
From the embedding of the Standard Model Effective Field Theory (SMEFT) in the more general Higgs Effective Field Theory (HEFT), we expose correlations among the coefficients of the latter that, if found to be violated in future data, would lead to the experimental falsification of the SMEFT framework. These are derived from the necessary symmetric point of HEFT and analyticity of the SMEFT Lagrangian that allows the construction of the SMEFT expansion, as laid out by other groups, and properties at that point of the Higgs-flare function $\mathcal{F}(h)$ coupling Goldstone and Higgs bosons, of the Higgs potential $V(h)$ and of the Higgs-top quark coupling function $\mathcal{G}(h)$.
\end{abstract}
%%%%%%%%%%%%%%%%%%%%%%%%%%%%%%%%%%%%%%%%%%%%%%%%%%%%%%%%%%%%%%%%%%%%%
%\maketitle
\end{frontmatter}

\section{Introduction}
Discovering new particles would entail the Standard Model (SM) being falsified in Popper's sense~\cite{Popper1934-POPLDF-3} and force us to extend it. Absent such discovery, 
the SM is still falsifiable upon finding new forces among the known particles. Because the SM has a characteristic energy scale of 100 GeV (the mass of the Higgs boson at $m_h=125$ GeV, as well as the vacuum constant $v=246$ GeV exemplify it), but no new particles below 1000 GeV have been found, there is a scale separation that begs the use of Effective Field Theory.

The popular SMEFT extension of the SM Electroweak Symmetry Breaking Sector (EWSBS) adds to it operators classified by their mass dimension,
\begin{equation}
    \mathcal{L}_{\rm SMEFT} =
    \mathcal{L}_{\rm SM} + 
    \sum_{n=5}^{\infty}
    \sum_i
    \frac{c_i^{(n)}}{\Lambda^{n-4}} \op_i^{(n)}(H) \ .
 \label{SMEFTL}
 \end{equation}
These operators $\op_i^{(n)}(H)$,  whose intensity is controlled by the  Wilson coefficients $c_i^{(n)}$ for a given reference scale $\Lambda$, are the potentials of those new forces being sought. 
A nonzero $c_i^{(n)}$ would signal departure from the SM, that then would need to be extended,
%\textcolor{blue}
{perhaps by new resonances~\cite{Dobado:2017lwg}}.

But it is easy to ask oneself how the whole framework of SMEFT can be tested. 
Effective theories include all the possible interactions that are compatible with the known particle content and symmetries believed to hold. Would it not be that any separation from the SM could be recast in SMEFT form? In that case, absent some new light particle, any phenomena could be described by adding an operator with a parameter to the SM. This is not so, as we will detail. 

The particle content of the electroweak symmetry breaking sector is packaged in a Higgs doublet field in the SM as in SMEFT
\begin{equation}
H = 
\frac{1}{\sqrt{2}} \begin{pmatrix} \varphi_1+i\varphi_2 \\   \varphi_0 + i\varphi_3
\end{pmatrix} 
= 
U(\boldsymbol{\omega}) \begin{pmatrix} 0 \\   (v+h_{\rm SMEFT})/\sqrt{2}
\end{pmatrix} \, ,
\end{equation}     
where the Cartesian coordinates $\varphi_a$ can be rearranged to the polar decomposition in terms the $\omega_i$ Goldstone bosons (which set the orientation of $H$ through the unitary matrix $U(\boldsymbol{\omega})$) and the radial coordinate $h_{\rm SMEFT}$ (with $|H|=(v+h_{\rm SMEFT})/\sqrt{2}$).

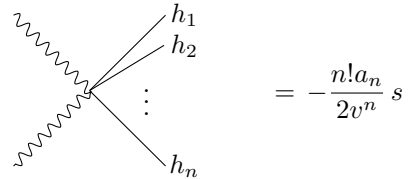
\begin{figure}[ht!]  %%%[ht!]
\centering
     \begin{tikzpicture}[scale=1]
     \draw[decoration={aspect=0, segment length=1.8mm, amplitude=0.7mm,coil},decorate] (-1,1) -- (0,0)-- (-1,-1);
     \draw[] (0,0)-- (1,1);
     \draw[] (1.25,1) node {$h_1$};
          \draw[] (1.25,-1) node {$h_n$};
     \draw[] (0,0)-- (.98,0.6);
          \draw[] (1.25,0.6) node {$h_2$};
     \draw[] (.75,-.15) node {.};
     \draw[] (.75,0.) node {.};
     \draw[] (.75,-0.3) node {.};
     \draw[] (0,0)-- (1,-1);
       \draw[] (3.25,0) node {$\displaystyle{  \, =\,  -\frac{n!a_n}{2v^n} \,  s }$};  
\end{tikzpicture}
\caption{\label{fig:Feynman}\small
The $\omega\omega\to nh$ processes can be the key to disentangling the nature of the EWSBS. They give direct access to the $a_i$ coefficients of the flare function $\mathcal{F}$, and hence to their correlations, as listed in Table~\ref{tab:correlations}.
Current experimental constraints do not really extend beyond $n=2$ and it will be challenging to examine higher coefficients. Better reconstruction techniques at the high-luminosity LHC run but especially a future high-energy collider (either hadronic or muonic) will hopefully improve the situation.
}
\end{figure}

\begin{figure}[!t]
    \centering
    \includegraphics[width=\columnwidth]{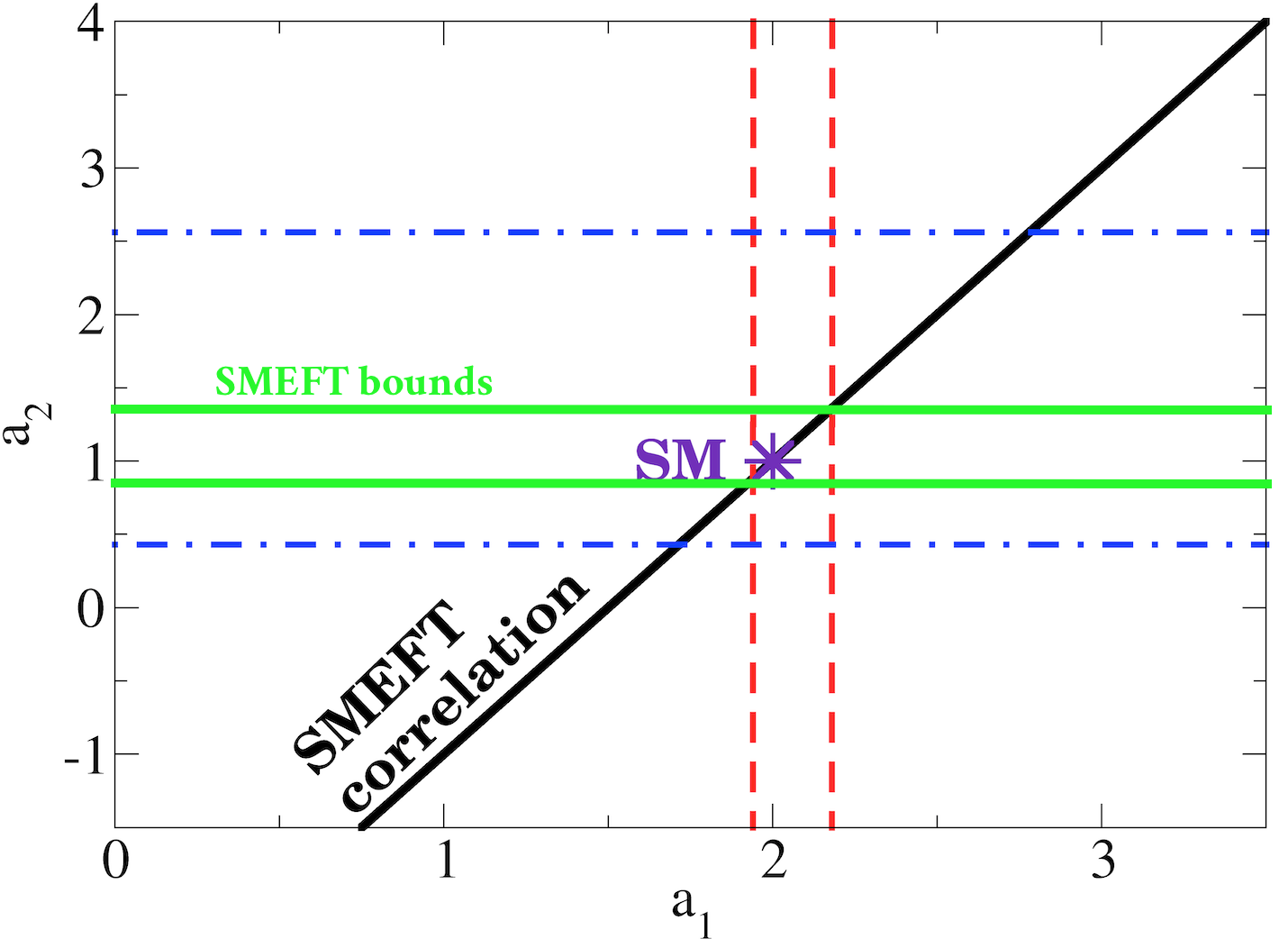}
    \caption{\small The correlation $a_2=2a_1-3$ that SMEFT predicts at order $1/\Lambda^2$ is plotted against the current 95\% confidence intervals for these two HEFT parameters~\cite{ATLAS:2020qdt,CMS:2022cpr}. The dashed blue lines correspond to the direct experimental bounds prior to this work, while the solid green lines represent the bounds in Table~\ref{tab:correlations} coming from our $1/\Lambda^2$ correlation analysis. }
    \label{fig:a1a2}
\end{figure}

%%%
An additional non-linear redefinition of $h_{\rm SMEFT}$ allows us to rearrange the SMEFT Lagrangian in the form of a  more general theory, HEFT: 
\begin{align}  \label{HEFTL}
{\cal L}_{\rm HEFT} = \frac{1}{2}\partial_\mu h_{\rm HEFT}\partial^\mu h_{\rm HEFT}-V(h_{\rm HEFT}) + \nonumber \\ 
+\frac{1}{2}\mathcal{F}(h_{\rm HEFT})
\partial_\mu \omega^i \partial^\mu \omega^j\!\left(\!\delta_{ij}\!+\!\frac{\omega^i\!\omega^j}{v^2\!-\!\boldsymbol{\omega}^2}\!\right)\ .
\end{align}
Of current focus therein is the flare function~\cite{Alonso:2016oah, Grinstein:2007iv} 
\begin{equation} \label{Fexpansion}
    {\mathcal F}(h_{\rm HEFT})=1+\sum_{n=1}^{\infty}{a_n}\Big(\frac{h_{\rm HEFT}}{v}\Big)^n \,,
\end{equation}
which amounts to a radial ``scale'' (think of $a(t)$ in a Fried-mann$-$Robertson$-$Walker cosmology) in the field space of the $(h,\omega_i)$ electroweak bosons (with $\omega_i$ analogous to the spatial coordinates).  
What we call attention to in this letter is that the Taylor-series coefficients of $\mathcal{F}$ as defined in Eq.~(\ref{Fexpansion}) must satisfy experimental correlations or constraints as given in Table~\ref{tab:correlations} similar to the ones in~\cite{Gomez-Ambrosio:2022qsi} 
%%%~\cite{Ambrosio:2022} 
{\it if SMEFT is a valid description}~\footnote{
Our advance since our previous publication in~\cite{Gomez-Ambrosio:2022qsi} consists of the new correlations among the HEFT 
$t$ quark-Yukawa sector coefficients, see Eq.~(\ref{eq:Yukawacorrelations})
and the improvements to the correlations among the Higgs self-interaction presented in Eq.~(\ref{eq:potentialfullcorrelations}) due to the $c_{H\Box}$ operator.
The results presented here have been reported to the QCHS conference resulting in a preliminary publication~\cite{Salas-Bernardez:2022hqv}.}. It is clear that an experimental program aimed at these correlations
via the key process to access $\mF$, $\omega\omega\to nh$ as sketched in Figure~\ref{fig:Feynman}, $mh\to nh$ to access $V(h_{\rm HEFT})$ (the Higgs potential), or $\bar{t}t\to nh$ to access the tree-level function modifying the Yukawa couplings, see Eq.~(\ref{eq:Gyukawa}), $\mathcal{G}(h)$, can test the validity of SMEFT itself, and not only its parameters. As an example, the SMEFT correlation among $a_1$ and $a_2$ is shown in Figure~\ref{fig:a1a2}.
\begin{table}[!t]  %%%[h]
%%% Table formatting %%%%%%%%% 
\setlength{\arrayrulewidth}{0.3mm} %thickness of table lines 
\setlength{\tabcolsep}{0.2cm}  %%%18pt} %separation of vertical lines to text 
\renewcommand{\arraystretch}{1.4} %space between rows 
% Table formatting %%%%%%%%%%%
    \caption{\small Correlations between the $a_i$ HEFT coefficients necessary for SMEFT to exist, at order $\Lambda^{-2}$ (first and second columns, with the numbers in the second consistent with 95\% confidence-level experimental bounds $a_1/2 \in [0.97,1.09]$~\cite{ATLAS:2020qdt}).
    The right column provides the corresponding numerical values at the next order~\cite{Gomez-Ambrosio:2022qsi}. 
    %%%\cite{Ambrosio:2022}. 
    They are quoted in terms  of $\Delta a_1:=a_1-2$ and $\Delta a_2:=a_2-1$, so that all  objects in the table vanish in the Standard Model, with all the equalities becoming $0=0$.    
    \label{tab:correlations}}
    \centering
    \begin{tabular}{|c|c|c|}\hline
        $\mathcal{O}(1/\Lambda^2)$ & $\mathcal{O}(1/\Lambda^2)$ & $\mathcal{O}(1/\Lambda^4)$ \\ \hline 
        $\Delta a_2=2\Delta a_1$ & $\Delta a_2\in [-0.12,0.36] $  &  \\
        $a_3=\frac{4}{3} \Delta a_1$ &  $a_3\in[-0.08,0.24]$  & $a_3\in [-3.1,1.7]$  \\  
        $a_4=\frac{1}{3} \Delta a_1$ &  $a_4\in [-0.02,0.06]    $ &  $a_4\in [-3.3,1.5]$ \\
        $a_5=0$ &  & $a_5 \in[-1.5,0.6]$ \\
        $a_6=0$ &  & $ a_6=a_5$ \\ 
        \hline
    \end{tabular}
\end{table}

There are several reasons why the experimental tests of those correlations need large energies and statistics, at the limit of what is possible today at the LHC and beyond.
%%%
First, the $\mathcal{F}$ function multiplies terms with derivatives of the Goldstone bosons $\partial_\mu \omega_i\to q_\mu \omega_i$ that yield couplings proportional to their four-momenta, and become more relevant at higher energies. 
%%%
Second, 
 the equivalence theorem~\cite{Veltman:1989ud,Dobado:1993dg} tells us that the scattering of longitudinal gauge bosons is related to scattering of Goldstone boson ($\omega_i \sim W^\pm_L,\ Z_L$): the EW gauging of the HEFT Lagrangian~(\ref{HEFTL}) leads to the $WW\to n h$ interaction $$\Delta\mathcal{L}_{\rm HEFT}^{WW\to n h} = \left(\frac{1}{2}m_Z^2 Z_\mu Z^\mu + m_W^2 W_\mu^+ W_\mu^-\right) \mathcal{F}(h_{\rm HEFT}),$$ which for longitudinal gauge bosons clearly dominates over the non-derivative interactions from $V$ only  at high energies~\cite{Dicus:1987ez,Kallianpur:1988cs}.  
And third, an increasing number of Higgs bosons  (necessary to access each $h^n$ order of $\mathcal{F}$, the Higgs-flare function) requires an ample phase space and, thus, high energy.

\section{Correlations in HEFT parameters induced by assuming SMEFT's validity}

As we shortly show after
Eq.~(\ref{changeofvariable}) below, the correlations mentioned above arise from the need for
consistency of the SMEFT formulation when a change of variable $h_{\rm HEFT}\to h_{\rm SMEFT}$ is performed. 
This change affects any other piece of the Lagrangian involving the Higgs bosons, such as the Yukawa couplings to fermions, saliently the top quark, or the interactions among Higgs bosons themselves (both of which we examine here), as well as couplings to transversal gauge bosons (that we leave for future works). 

\begin{figure}[!t]
    \centering
    \includegraphics[width=\columnwidth]{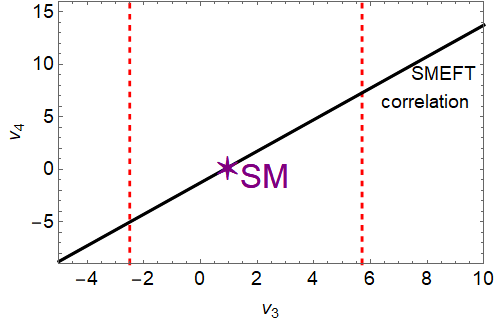}
    \caption{\small The correlation $v_4=\frac{3}{2} v_3 -\frac{5}{4} -\frac{1}{6}\Delta a_1$ that SMEFT predicts at $\mathcal{O}(1/\Lambda^2)$ is plotted making use of current 95\% confidence interval for $v_3\in[-2.5,5.7]$~\cite{ATLAS:2021jki}. The experimental $a_1$ uncertainty~\cite{ATLAS:2020qdt,CMS:2022cpr},  $a_1/2\in[0.97,1.09]$, is numerically negligible and allows to  predict a SMEFT band given by the solid black line. An experimental measurement for $v_4$ is still missing. }
    \label{fig:a1a22}
\end{figure}

The much discussed $V(H)$ Higgs-potential, experimentally accessible at ``low'' $\sqrt{s}$ because it contains no derivative couplings,
\begin{equation}
{\mathcal{L}}_{\rm SM} = |\partial H|^2 -   \underbrace{\left( \mu^2 |H|^2 +\lambda|H|^4\right)}_{V(H)} \, ,
\end{equation}
 acquires in HEFT additional non-renormalizable couplings  organized in a power-series expansion
\begin{align}\label{expandV}
    V_{\rm HEFT}= \frac{m_h^2 v^2}{2}  \Bigg[&   \left(\frac{h_{\rm HEFT}}{v}\right)^2 +  v_3   \left(\frac{h_{\rm HEFT}}{v}\right)^3 \nonumber+ \\  &+ v_4    \left(\frac{h_{\rm HEFT}}{v}\right)^4 + \dots \Bigg]\,,
\end{align}
with $v_3=1$, $v_4=1/4$ and $v_{n\geq 5}=0$ in the SM. 
Its coefficients also need to satisfy constraints that are exposed in Table~\ref{tab:corV} and Figure~\ref{fig:a1a22} if and when SMEFT applies.

\begin{figure}[!t]
    \centering
    \includegraphics[width=\columnwidth]{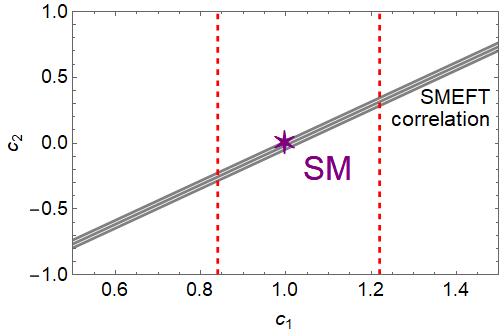}
    \caption{\small 
     The correlation , $c_2=\frac{3}{2} (c_1 -1) -\frac{1}{4}\Delta a_1$, that SMEFT predicts at $\mathcal{O}(1/\Lambda^2)$ is plotted making use of current 95\% confidence interval for the top Yukawa coupling $c_1\in[0.84,1.22]$\cite{deBlas:2018tjm} (dashed red lines). The experimental $a_1$ uncertainty~\cite{ATLAS:2020qdt,CMS:2022cpr},  $a_1/2\in[0.97,1.09]$ (at 95\% CL), is reflected in the width of the gray band with the SMEFT correlation. An experimental determination of the $ t\bar{t}\to hh$ coupling $c_2$ is still missing.   
   However, should it be measured, a test of SMEFT is possible by comparing to $c_2\in[-0.27,0.35]$  (from linearly adding uncertainties with the existing data and analyses).}
    \label{fig:c1c2}
\end{figure}

Similarly, the SM piece coupling the top quark to the Higgs boson is extended in HEFT~\cite{Castillo:2016erh} by a multiplicative function $\mathcal{G}(h)$ 
\begin{equation}\label{eq:Gyukawa}
\mathcal{L}_Y= -\mathcal{G}(h) M_t \bar{t} t 
%%%\left( 1-\frac{2\omega^2}{v^2}\right) \;,
\sqrt{1-\frac{\omega^2}{v^2}}\; ,
\end{equation}
with a Taylor expansion around the physical $h=0$ vacuum given by
\begin{equation} \label{Yukawa}
    \mathcal{G}(h_{\rm HEFT})= 1 + c_1 \frac{h_{\rm HEFT}}{v} + c_2 \left( \frac{h_{\rm HEFT}}{v} \right)^2+\dots \end{equation}  
    (with $c_1=1$, $c_{i\geq 2}= 0$ 
in the Standard Model).
The correlations among these coefficients induced by SMEFT at order $1/\Lambda^2$  are then again given in Table~\ref{tab:corV} and Figure~\ref{fig:c1c2}.

Let us then see, very briefly, how the various correlations come about. Alternatively to relying on the powerful geometric methods of~\cite{Alonso:2016oah,Alonso:2015fsp,Alonso:2016btr,Alonso:2021rac,Alonso:2022ffe} we use the more pedestrian coordinate-dependent approach, more familiar to phenomenologists working on LHC physics.
The goal is to see when is it possible to cast  
Eq.~(\ref{HEFTL}) into the specific SMEFT one, Eq.~(\ref{SMEFTL}).
This we write as 
\begin{align}
 \mathcal{L}_{\rm SMEFT}= |\partial H|^2-V(|H|^2)
+\frac{1}{2} B(|H|^2)(\partial (|H|^2))^2 +\dots  
\label{eq:SMEFTL-kin+V+B}
\end{align}
where the  non-derivative and derivative terms, respectively given by $V$ and $B$, 
collect typical SMEFT operators 
(think of them as expressed in the Warsaw  
basis). Note that we have only kept the partial derivative part of the SM Higgs doublet kinetic term $|DH|^2$ in the r.h.s. of~(\ref{eq:SMEFTL-kin+V+B}), as we are considering the equivalence theorem and we are focused on the scalar sector of the theory.  
At the lowest order correction, $1/\Lambda^2$, the relevant dimension--6 SMEFT operators for our analysis are, 
\begin{equation}\label{operatorsSMEFT}
  \op_H := (H^\dagger H)^3  \, , \ \ \ 
  \op_{H \Box} := (H^\dagger H) \Box (H^\dagger H)\ .
\end{equation} 
 There are also other operators, such as, e.g., 
$  \op_{HD} = (H^\dagger D_{\mu} H)^*  (H^\dagger D^{\mu} H) $, but they break custodial symmetry, and LEP studies suggest that the $SU(2)\times SU(2)\to SU(2)$ electroweak symmetry breaking mechanism is the appropriate pattern, leaving the residual custodial $SU(2)$ as a good approximate global symmetry of the scalar sector.  
The additional $A(H)$ structure pointed out in~\cite{Cohen:2020xca} for Lagrangian~(\ref{eq:SMEFTL-kin+V+B}) can be eliminated through partial integration and the use of the equations of motion~\cite{Gomez-Ambrosio:2022qsi}.

\begin{table}[!t]
\setlength{\arrayrulewidth}{0.3mm} 
\setlength{\tabcolsep}{0.2cm}  
\renewcommand{\arraystretch}{1.4} 
\caption{\small Correlations among the coefficients $\Delta v_3:=v_3-1$, $\Delta v_4:=v_4-1/4$, $v_5$ and $v_6$ of the HEFT Higgs potential expansion in Eq.~(\ref{expandV}) that need to hold, at $\mO(1/\Lambda^2)$, if SMEFT is a valid description of the electroweak sector.
Based  on the current bound $\Delta v_3\in[-2.5,5.7]$ in Ref.~\cite{ATLAS:2021jki}, $\mO(1/\Lambda^2)$ SMEFT predicts the coefficient intervals in the last column, testable in few-Higgs final states. A coupling 
$c_{H\Box}\neq 0$ induces the correction $\Delta a_1\propto c_{H\Box}$, nevertheless numerically negligible since $v_3$ experimental uncertainties much exceed those of $a_1$. Likewise, we include the leading correlations for the Yukawa $\mathcal{G}(h)$ function of Eq.~(\ref{Yukawa}), constraining $c_2$ and $c_3$ by $c_1$ and $a_1$ (from the correction to the value of the symmetric point $h_*$). We make use of current 95\% confidence interval for the top Yukawa coupling $c_1\in[0.84,1.22]$~\cite{deBlas:2018tjm}.}   
\label{tab:corV}
\begin{center}
 \begin{tabular}{|c|c|} \hline 
        $\Delta v_4=\frac{3}{2}\Delta v_3  -\frac{1}{6}\Delta a_1$ & $\Delta v_4\in[-3.8,8.6]$\\[2ex]
        $ v_5=6v_6=\frac{3}{4}\Delta v_3 -\frac{1}{8}\Delta a_1 $ 
        & $v_5=6 v_6 \in[-1.9,4.3]$ \\ \hline 
        $c_2
       = 3 c_3  
        =\frac{3}{2} (c_1 -1) -\frac{1}{4}\Delta a_1$ & $c_2
        = 3 c_3  
        \in[-0.27,0.35]$\\
%%%        $c_3=\frac{1}{2} (c_1 -1) -\frac{1}{12}\Delta a_1$ & $c_3\in[-0.09,0.12]$        
%%%        \\[2ex] 
%%%           $v_6=\frac{1}{6}v_5$  %%%\frac{1}{8}\Delta v_3  -\frac{1}{48}\Delta a_1$ 
%%%           &  $v_6=\frac{1}{6}v_5$   %%%$v_6\in[-0.3,0.7]$ 
%%%           \\
         \hline    
    \end{tabular}
\end{center}  
\end{table}
 \newpage
To proceed, we need to perform the following conversion to pass from SMEFT to HEFT and viceversa:  
 \begin{align}
 &|\partial H|^2 +  
\frac{1}{2} B(|H|^2)(\partial (|H|^2))^2 \longleftrightarrow    \nonumber \\
&\frac{1}{2} (\partial h_{\rm HEFT})^2 + \frac{v^2}{4} \mathcal{F}(h_{\rm HEFT}) \,   {\rm Tr}\{ \partial_\mu U^\dagger \partial^\mu U\}
\,,
\end{align}  
The change from SMEFT to HEFT is straightforward and always possible, with the canonical, nonlinear change of variables given in differential form as
\begin{equation}  \label{changeofvariable}  
dh_{\rm HEFT}\, =\, \sqrt{1+(v+h_{\rm SMEFT})^2 B(h_{\rm SMEFT})}\,\, dh_{\rm SMEFT} \,,
\end{equation}
where the flare-function is provided by the relation 
\begin{equation}
\mathcal{F}(h_{\rm HEFT}) \,=\, \left(1+h_{\rm SMEFT}/v\right)^2\, .
\end{equation}  
However,  
the reverse conversion from HEFT to SMEFT, 
\begin{equation}
h_{\rm HEFT}   \,=\, \mathcal{F}^{-1}\left((1+h_{\rm SMEFT}/v)^2\right) \,,
\end{equation}
runs into difficulty. 
This is because of the need to reconstruct squared operators of the Higgs doublet field $H$ that is the basis of SMEFT, such as
\begin{eqnarray}
|H|^2 &=& \frac{(v+h_{\rm SMEFT})^2}{2}\, ,
\nonumber \\
(\partial|H|^2)^2 &=& (v+h_{\rm SMEFT})^2 \,   (\partial h_{\rm SMEFT})^2 \nonumber \\ &=&\, 2 |H|^2 \,   
(\partial h_{\rm SMEFT})^2 \, .
\end{eqnarray} 
The extra $|H|^2$ on the right hand side of the second equation 
ends in a denominator

\begin{align}
&\mathcal{L}_{\rm SMEFT} =  \underbrace{ |\partial H|^2}_{= \mathcal{L}_{\rm SM}} \quad +\quad %%%\nonumber  
\label{eq:HEFT2SMEFT}
\\ \nonumber 
&+\underbrace{ \frac{1}{2} \bigg[  \frac{8|H|^2}{v^2}\bigg(   (\mathcal{F}^{-1})'\left(2| H |^2/v^2\right)  \bigg)^2 \,\,\,-\,\,\, 1\bigg] \, \frac{(\partial| H |^2)^2}{2| H |^2}  }_{=\Delta \mathcal{L}_{\rm BSM}}\, .
\end{align}
%\\ no?$>>>>$  \\  
As SMEFT is assumed to have the analytical power expansion
in Eq.~(\ref{SMEFTL}), such singularity precludes its existence and needs to be cancelled by the preceding bracket in the second line of Eq.~(\ref{eq:HEFT2SMEFT}). 

The result is the same as that obtained by geometric methods~\cite{Cohen:2020xca}, there must be a double zero of $\mathcal{F}$, a symmetric point with respect to the global $SU(2)\times SU(2)$ group so that the SMEFT expansion can be performed.  Furthermore,
analyticity requires that all its odd derivatives vanish at the symmetric point.

The particular case of the SM is given by $\mathcal{F}=(1+h_{\rm SMEFT}/v)^2 $. As already pointed out,
at higher orders in $h/v$, the existence of SMEFT requires that the odd derivatives of $\mathcal{F}$ at the symmetric point $h_\ast$ vanish. 

The correlations from Table~\ref{tab:correlations} can then be obtained by matching the Taylor expansion of $\mathcal{F}$ around such symmetric point 
$ h_{\rm HEFT}=  h_\ast$ with the expansion around our physical vacuum $h_{\rm HEFT}=0$.
Instead of that matching, one can also obtain the correlations by eliminating the SMEFT Wilson coefficients order by order. For example, at $\mO(1/\Lambda^2)$ there is only one operator, $\mathcal{O}_{H\Box}$,  in Eq.~(\ref{operatorsSMEFT}), that controls all the HEFT coefficients of $\mathcal{F}$: 
\begin{align}
&& %%%\;\;\;
a_1 = 2a=2\left(1 + v^2\frac{c_{H\Box}}{\Lambda^2}\right)\,, \quad 
%%%&,& \;\;\;
a_2 = b=1+{4v^2}\frac{c_{H\Box}}{\Lambda^2} \,, 
\nonumber \\ 
&& %%%\;\;\;
a_3 = \frac{8v^2}{3}\frac{c_{H\Box}}{\Lambda^2}\,, \qquad 
%%%&,& \;\;\;
a_4 = \frac{2v^2}{3}\frac{c_{H\Box}}{\Lambda^2}\,,\qquad 
a_{n\geq 5} = 0 \, .
\end{align}
The elimination, by substitution, of this $c_{H\Box}$ coefficient from the HEFT $a_i$'s yields the $1/\Lambda^2$ correlations of the second column of Table~\ref{tab:correlations}. Proceeding to the next $1/\Lambda^4$ order in the SMEFT expansion 
brings in the Wilson coefficient $c^{(8)}_{H\Box}$. Hence, 
one can likewise extract the weaker $1/\Lambda^4$ correlations among the HEFT parameters.
The potential $V(h_{\rm HEFT})$ is in turn also affected by $\mathcal{O}_H$,
\begin{align}
&& 
v_3 =1 +  \frac{3v^2 c_{H\Box}}{\Lambda^2} +\epsilon_{c_H}
\,, \,\,\, 
v_4 = \frac{1}{4} +  \frac{25v^2c_{H\Box} }{6\Lambda^2} +\frac{3}{2}\epsilon_{c_H}
\,,
\nonumber\\
&& 
v_5 =    \frac{2v^2c_{H\Box} }{\Lambda^2} +\frac{3}{4}\epsilon_{c_H}
\,,\,\,\, 
v_6 =  \frac{v^2c_{H\Box} }{3\Lambda^2} +\frac{1}{8}\epsilon_{c_H}
\,,
\nonumber\\
&& 
v_{n\geq 7} = 0 \,, \label{eq:potentialfullcorrelations}\qquad\qquad\qquad 
\end{align}
with  $m_h^2=\, -2\mu^2 \left(1+\frac{2 c_{H\Box}v^2}{\Lambda^2}+ \frac{3}{4}\epsilon_{c_H}\right)$, $2\langle |H|^2\rangle =v^2= -\frac{\mu^2}{\lambda}\left(1-\frac{3}{4}\epsilon_{c_H}\right) $ and $\epsilon_{c_H}= - \frac{2 v^4 c_H}{m_h^2\Lambda^2}= \frac{\mu^2 c_H}{\lambda^2  \Lambda^2}$. 
%%%, $\Delta a_1=  \frac{2c_{H\Box}v^2}{\Lambda^2}$. 

Also, the $c_i$ in $\mathcal{G}(h_{\rm HEFT})$ modifying the Yukawa coupling receive analogous contributions from both SMEFT coefficients $c_{H\square}$ and $c_{uH}$ in standard notation, the second alternatively named $c_{tH^+}$ in~\cite{Brod:2022bww}. The correction
\begin{equation}\label{eq:Yukawacorrelations}
c_1 = 1 -\frac{v^3}{\sqrt{2}m_t} \frac{c_{tH^+}}{\Lambda^2} 
  +\frac{c_{H\square}v^2}{\Lambda^2}+\mO(1/\Lambda^4) 
  \end{equation}
can be carried on to the higher coefficients using the relations in Table~\ref{tab:corV} 
(with $\Delta a_1=2 c_{H\square} v^2/\Lambda^2+\mO(1/\Lambda^4)$).

%%%%%%%%%%%%%%%%%%%%%%%%%%%%%%%%%%%%%%%%%
\newpage
\section{Conclusions}

%%%\color{blue}{Brod:2022bww
Various authors, see {\it e.g.}~\cite{Brivio:2016fzo} have pointed out to differences between the  SMEFT and HEFT formulations~\cite{Dobado:2019fxe}.
%
%%%\color{black}
For example, in SMEFT the Goldstone
 $\omega_i$  and Higgs  $h_{\rm SMEFT}$ bosons are arranged in a left-$SU(2)$ doublet while in HEFT
$h_{\rm HEFT}$ is an $SU(2)\otimes SU(2)$ singlet, independent of the Goldstone triplet $\omega_i$.  
Also, in SMEFT the Higgs field always appears  in the combination $ (h_{\rm SMEFT} + v)$ 
and thus, HEFT deploys more independent higher-dimension effective operators  (in exchange, it is less model dependent).
This means that SMEFT is natural when $h_{\rm SMEFT}$ is a fundamental field while HEFT is typical for composite models of the EWSBS (such as those with $h_{\rm HEFT}$ as a Goldstone boson).
And finally, the counting of SMEFT is based in a cutoff $\Lambda$ expansion taking the canonical operator dimensions, $\mathcal{O}(d)/\Lambda^{d-4}$ (independently of $N_{\rm loops}$) whereas HEFT is a derivative expansion (independently of $N_{\rm particles})$ like the older Electroweak Chiral Lagrangian, with $\mathcal{F}(h)$ inserted in the derivative Goldstone term.

Nevertheless, a lot of this is cosmetic and can be reorganized by changing variables $h_{\rm SMEFT}\leftrightarrow h_{\rm HEFT}$. What is key is the San Diego criterion~\cite{Alonso:2016oah,Alonso:2015fsp}:  
$\mathcal{F}(h_{\rm HEFT})$ must have a point $h_\ast$ symmetric under the global $SU(2)\times SU(2)$  group  and due to its existence and convergence in the $h$ field space, SMEFT is deployable if and only if  (which is a statement about the HEFT Lagrangian)
\begin{itemize}
\item   $\exists h_\ast \in\mathbb{R}$ where $\mathcal{F} (h_\ast)=0$, and 
\item  Because of the need for $\mathcal{L}_{\rm SMEFT}$ analyticity, $\mathcal{F}$ is analytic between our vacuum $h=0$ and $h_\ast$, particularly around $h_\ast$. Moreover its odd derivatives vanish.
\end{itemize}
We have presented new relations that implement this criterion at $\mathcal{O}(1/\Lambda^2)$ and $\mathcal{O}(1/\Lambda^4)$ in the $1/\Lambda$ counting; more precision is unnecessary until (if) separations from the Standard Model are found. Then only, with the scale $\Lambda$ at hand out of separations of EFT coefficients from the SM, can we decide how relevant the corrections due to the higher orders are expected to be, and whether further work is warranted.

Among the three types of correlations that we have presented in tables~\ref{tab:correlations} and~\ref{tab:corV}, those for the coefficients of $\mathcal{F}$ are more interesting for large values of the energy $\sqrt{s}\gg m_h\sim m_W \sim m_Z$ whereas those for $V$ and $\mathcal{G}$, that do not involve Goldstone bosons, are therefore of greater interest at low energies, when
the potential competes with the derivative operators on equal ground, as $\sqrt{s} \sim m_i$.

In conclusion, we have newly translated these conditions into correlations among HEFT coefficients 
whose violation falsifies SMEFT. 
Moreover, since many extensions of the Standard Model incorporating supersymmetry, supergravity, or other possibilities, can be cast as a SMEFT, they can be likewise simultaneously falsified.

For the time being, 
%%%\textcolor{blue}
{no separations from the SM have been found~\cite{Eboli:2021unw} and} one can only infer direct experimental bounds on  the first  terms, $a_1$ and, perhaps, $a_2$, so we have to wait for data with a larger number of Higgs bosons before assessing them. But when this will be done, the correlations will allow to falsify SMEFT in experiment
even without new particles. We believe that this possibility improves the standing of SMEFT as a scientific theory.

%%%%%%%%%%%%%%%%%%%%%%%%%%%%%%%%%%%%%%%%%%%%%%%%%%%%%%%%%%%%%%%%%%%%%%%%%%%%%%%%%%%%%%
\vspace{.4cm}
\section*{Acknowledgments}
Supported by spanish MICINN PID2019-108655GB-I00/AEI/10.13039/501100011033 grant, and Universidad Complutense de Madrid under research group 910309 and the IPARCOS institute; 
ERC Starting Grant REINVENT-714788; UCM CT42/18-CT43/18;
the Fondazione Cariplo and Regione Lombardia, grant 2017-2070: and  by Grant DataSMEFT23 (EUNextGeneration - PNRR - DM 247 08/22).

\newpage
%%%%%%%%%%%%%%%%%%%%%%%%%%% 
%%%\color{blue}
\bibliography{externalbibforSMEFTtoHEFT}

\begin{thebibliography}{10}
\expandafter\ifx\csname url\endcsname\relax
  \def\url#1{\texttt{#1}}\fi
\expandafter\ifx\csname urlprefix\endcsname\relax\def\urlprefix{URL }\fi
\expandafter\ifx\csname href\endcsname\relax
  \def\href#1#2{#2} \def\path#1{#1}\fi

\bibitem{Popper1934-POPLDF-3}
K.~Popper, Logik der forschung, Erkenntnis 5~(1) (1934) 290--294.

\bibitem{Dobado:2017lwg}
A.~Dobado, F.~J. Llanes-Estrada, J.~J. Sanz-Cillero, {Resonant production of Wh
  and Zh at the LHC}, JHEP 03 (2018) 159.
\newblock \href {http://arxiv.org/abs/1711.10310} {\path{arXiv:1711.10310}},
  \href {https://doi.org/10.1007/JHEP03(2018)159}
  {\path{doi:10.1007/JHEP03(2018)159}}.

\bibitem{Alonso:2016oah}
R.~Alonso, E.~E. Jenkins, A.~V. Manohar, {Geometry of the Scalar Sector}, JHEP
  08 (2016) 101.
\newblock \href {http://arxiv.org/abs/1605.03602} {\path{arXiv:1605.03602}},
  \href {https://doi.org/10.1007/JHEP08(2016)101}
  {\path{doi:10.1007/JHEP08(2016)101}}.

\bibitem{Grinstein:2007iv}
B.~Grinstein, M.~Trott, {A Higgs-Higgs bound state due to new physics at a
  TeV}, Phys. Rev. D 76 (2007) 073002.
\newblock \href {http://arxiv.org/abs/0704.1505} {\path{arXiv:0704.1505}},
  \href {https://doi.org/10.1103/PhysRevD.76.073002}
  {\path{doi:10.1103/PhysRevD.76.073002}}.

\bibitem{Gomez-Ambrosio:2022qsi}
R.~G\'omez-Ambrosio, F.~J. Llanes-Estrada, A.~Salas-Bern\'ardez, J.~J.
  Sanz-Cillero, {Distinguishing electroweak EFTs with $W_L W_L \to n$ x $h$},
  Phys. Rev. D 106~(5) (2022) 053004.
\newblock \href {https://doi.org/10.1103/PhysRevD.106.053004}
  {\path{doi:10.1103/PhysRevD.106.053004}}.

\bibitem{Salas-Bernardez:2022hqv}
A.~Salas-Bernardez, J.~J. Sanz-Cillero, F.~J. Llanes-Estrada,
  R.~Gomez-Ambrosio, {SMEFT as a slice of HEFT\textquoteright{}s parameter
  space}, EPJ Web Conf. 274 (2022) 08013.
\newblock \href {http://arxiv.org/abs/2211.09605} {\path{arXiv:2211.09605}},
  \href {https://doi.org/10.1051/epjconf/202227408013}
  {\path{doi:10.1051/epjconf/202227408013}}.

\bibitem{ATLAS:2020qdt}
{A combination of measurements of Higgs boson production and decay using up to
  $139$ fb$^{-1}$ of proton--proton collision data at $\sqrt{s}=$ 13 TeV
  collected with the ATLAS experiment}, ATLAS-CONF-2020-027, communication by
  the ATLAS coll. to 40th ICHEP, Prague (8 2020).

\bibitem{CMS:2022cpr}
A.~Tumasyan, et~al., {Search for Higgs Boson Pair Production in the Four b
  Quark Final State in Proton-Proton Collisions at $\sqrt s$=13\,\,TeV}, Phys.
  Rev. Lett. 129~(8) (2022) 081802.
\newblock \href {http://arxiv.org/abs/2202.09617} {\path{arXiv:2202.09617}},
  \href {https://doi.org/10.1103/PhysRevLett.129.081802}
  {\path{doi:10.1103/PhysRevLett.129.081802}}.

\bibitem{Veltman:1989ud}
H.~G.~J. Veltman, {The Equivalence Theorem}, Phys. Rev. D 41 (1990) 2294.
\newblock \href {https://doi.org/10.1103/PhysRevD.41.2294}
  {\path{doi:10.1103/PhysRevD.41.2294}}.

\bibitem{Dobado:1993dg}
A.~Dobado, J.~R. Pel\'aez, {On The Equivalence theorem in the chiral
  perturbation theory description of the symmetry breaking sector of the
  standard model}, Nucl. Phys. B 425 (1994) 110--136, [Erratum: Nucl.Phys.B
  434, 475--475 (1995)].
\newblock \href {http://arxiv.org/abs/hep-ph/9401202}
  {\path{arXiv:hep-ph/9401202}}, \href
  {https://doi.org/10.1016/0550-3213(94)90174-0}
  {\path{doi:10.1016/0550-3213(94)90174-0}}.

\bibitem{Dicus:1987ez}
D.~A. Dicus, K.~J. Kallianpur, S.~S.~D. Willenbrock, {Higgs Boson Pair
  Production in the Effective $W$ Approximation}, Phys. Lett. B 200 (1988)
  187--192.
\newblock \href {https://doi.org/10.1016/0370-2693(88)91134-3}
  {\path{doi:10.1016/0370-2693(88)91134-3}}.

\bibitem{Kallianpur:1988cs}
K.~J. Kallianpur, {Pair Production of Higgs Bosons via Heavy Quark
  Annihilation}, Phys. Lett. B 215 (1988) 392--396.
\newblock \href {https://doi.org/10.1016/0370-2693(88)91454-2}
  {\path{doi:10.1016/0370-2693(88)91454-2}}.

\bibitem{Castillo:2016erh}
A.~Castillo, R.~L. Delgado, A.~Dobado, F.~J. Llanes-Estrada,
  {Top\textendash{}antitop production from $W^+_L W^-_L$ and $Z_L Z_L$
  scattering under a strongly interacting symmetry-breaking sector}, Eur. Phys.
  J. C 77~(7) (2017) 436.
\newblock \href {http://arxiv.org/abs/1607.01158} {\path{arXiv:1607.01158}},
  \href {https://doi.org/10.1140/epjc/s10052-017-4991-6}
  {\path{doi:10.1140/epjc/s10052-017-4991-6}}.

\bibitem{Alonso:2015fsp}
R.~Alonso, E.~E. Jenkins, A.~V. Manohar, {A Geometric Formulation of Higgs
  Effective Field Theory: Measuring the Curvature of Scalar Field Space}, Phys.
  Lett. B 754 (2016) 335--342.
\newblock \href {http://arxiv.org/abs/1511.00724} {\path{arXiv:1511.00724}},
  \href {https://doi.org/10.1016/j.physletb.2016.01.041}
  {\path{doi:10.1016/j.physletb.2016.01.041}}.

\bibitem{Alonso:2016btr}
R.~Alonso, E.~E. Jenkins, A.~V. Manohar, {Sigma Models with Negative
  Curvature}, Phys. Lett. B 756 (2016) 358--364.
\newblock \href {http://arxiv.org/abs/1602.00706} {\path{arXiv:1602.00706}},
  \href {https://doi.org/10.1016/j.physletb.2016.03.032}
  {\path{doi:10.1016/j.physletb.2016.03.032}}.

\bibitem{Alonso:2021rac}
R.~Alonso, M.~West, {Roads to the Standard Model}, Phys. Rev. D 105~(9) (2022)
  096028.
\newblock \href {http://arxiv.org/abs/2109.13290} {\path{arXiv:2109.13290}},
  \href {https://doi.org/10.1103/PhysRevD.105.096028}
  {\path{doi:10.1103/PhysRevD.105.096028}}.

\bibitem{Alonso:2022ffe}
R.~Alonso, M.~West, {On the effective action for scalars in a general manifold
  to any loop order} (7 2022).
\newblock \href {http://arxiv.org/abs/2207.02050} {\path{arXiv:2207.02050}}.

\bibitem{Cohen:2020xca}
T.~Cohen, N.~Craig, X.~Lu, D.~Sutherland, {Is SMEFT Enough?}, JHEP 03 (2021)
  237.
\newblock \href {http://arxiv.org/abs/2008.08597} {\path{arXiv:2008.08597}},
  \href {https://doi.org/10.1007/JHEP03(2021)237}
  {\path{doi:10.1007/JHEP03(2021)237}}.

\bibitem{ATLAS:2021jki}
{Search for Higgs boson pair production in the two bottom quarks plus two
  photons final state in $pp$ collisions at $\sqrt{s}=13$ TeV with the ATLAS
  detector}, ATLAS-CONF-2021-016, contribution of the ATLAS coll. to the 55th
  Rencontres de Moriond (3 2021).

\bibitem{deBlas:2018tjm}
J.~de~Blas, O.~Eberhardt, C.~Krause, {Current and Future Constraints on Higgs
  Couplings in the Nonlinear Effective Theory}, JHEP 07 (2018) 048.
\newblock \href {http://arxiv.org/abs/1803.00939} {\path{arXiv:1803.00939}},
  \href {https://doi.org/10.1007/JHEP07(2018)048}
  {\path{doi:10.1007/JHEP07(2018)048}}.

\bibitem{Brod:2022bww}
J.~Brod, J.~M. Cornell, D.~Skodras, E.~Stamou, {Global constraints on Yukawa
  operators in the standard model effective theory}, JHEP 08 (2022) 294.
\newblock \href {http://arxiv.org/abs/2203.03736} {\path{arXiv:2203.03736}},
  \href {https://doi.org/10.1007/JHEP08(2022)294}
  {\path{doi:10.1007/JHEP08(2022)294}}.

\bibitem{Brivio:2016fzo}
I.~Brivio, J.~Gonzalez-Fraile, M.~C. Gonzalez-Garcia, L.~Merlo, {The complete
  HEFT Lagrangian after the LHC Run I}, Eur. Phys. J. C 76~(7) (2016) 416.
\newblock \href {http://arxiv.org/abs/1604.06801} {\path{arXiv:1604.06801}},
  \href {https://doi.org/10.1140/epjc/s10052-016-4211-9}
  {\path{doi:10.1140/epjc/s10052-016-4211-9}}.

\bibitem{Dobado:2019fxe}
A.~Dobado, D.~Espriu, {Strongly coupled theories beyond the Standard Model},
  Prog. Part. Nucl. Phys. 115 (2020) 103813.
\newblock \href {http://arxiv.org/abs/1911.06844} {\path{arXiv:1911.06844}},
  \href {https://doi.org/10.1016/j.ppnp.2020.103813}
  {\path{doi:10.1016/j.ppnp.2020.103813}}.

\bibitem{Eboli:2021unw}
O.~J.~P. Eboli, M.~C. Gonzalez-Garcia, M.~Martines, {Electroweak Higgs
  effective field theory after LHC run 2}, Phys. Rev. D 105~(5) (2022) 053003.
\newblock \href {http://arxiv.org/abs/2112.11468} {\path{arXiv:2112.11468}},
  \href {https://doi.org/10.1103/PhysRevD.105.053003}
  {\path{doi:10.1103/PhysRevD.105.053003}}.

\end{thebibliography}
%%%\color{black}
%%%%%%%%%%%%%%%%%%%%%%%%%%%

\end{document}